# The Thick-COBRA: a New Gaseous Electron Multiplier for Radiation Detectors

**F. D. Amaro** [a*], **C. Santos** [b], **J. F. C. A. Veloso** [b], **A. Breskin** [c], **R. Chechik** [c], **J. M. F. dos Santos** [a]

[a] *Physics Department, University of Coimbra*
  *Coimbra, Portugal*

[b] *Physics Department, University of Aveiro*
  *Aveiro, Portugal*

[c] *Department of Particle Physics, The Weizmann Institute of Science*
  *Rehovot, Israel*

  *E-mail*: famaro@gian.fis.uc.pt

ABSTRACT: The operation principle and preliminary results of a novel gas-avalanche patterned hole electron multiplier, the Thick-COBRA (THCOBRA), are presented. This micro-hole structure is derived from the THGEM and MHSP. Sub-millimeter diameter holes are mechanically drilled in a thin G10 plate, Cu-clad on both faces; on one of the faces the Cu is etched to produce additional anode strips winding between circular cathode strips. Primary avalanches occurring within the holes are followed by additional ones at the anode-strips vicinity. Gains in excess of $5*10^4$ were reached with 22.1 x-rays in Ar, Ne and Ar-10%$CH_4$, with 12.2 % FWHM energy resolution in Ar-10%$CH_4$. Higher gains were measured with single photoelectrons. This robust multiplier may have numerous potential applications.

# Contents

## 1. INTRODUCTION

We present the first results obtained with a new gaseous electron multiplier, the Thick COBRA (THCOBRA). This device is an extension of the Micro Hole and Strip Plate (MHSP) [1] to a thicker, more robust electron multiplier and combines a patterned electrode structure on top of a thick-GEM (THGEM )[2][3].

    The MHSP consists of a thin (50 micron) metalized insulator with a fine hole-pattern (typically 50 micron diameter, 200 micron pitch) and thin (10-30 micron) anode and cathode strips printed on one face. According to their polarization, the latter either multiply electrons from avalanches occurring within the holes or defocus avalanche ions, preventing them to return to the holes [4]. Since their introduction [1] MHSPs have been investigated in several applications, ranging from x-ray and neutron detection, ion back-flow reduction in gaseous detectors, 2-D x-ray imaging and detection of noble gas scintillation. The MHSP has proven to be a stable and reliable detector, operating at high charge gains in several gas mixtures, including high pressure pure noble gases, and under high radiation flux with an overall good energy resolution [1],[4]-[10]. It played a crucial role in avalanche-ion blocking in gaseous photomultipliers with bialkali photocathodes, sensitive to visible light, permitting for the first time the operation of such devices at charge gains of $10^5$ [11].

    Thick gaseous electron multipliers (THGEM) have been recently introduced [2] and immediately attracted attention due to their robustness and easiness of production. These sub-millimeter patterned structures, with 10-fold expanded dimensions compared to GEM [12], can be readily produced, in large quantities and at low cost, using standard printed circuit board (PCB) technology: mechanical drilling of sub-millimeter holes in printed circuit boards followed by chemical etching of the rims around each hole (for increased electrical robustness [3]). The electric field within the holes reaches values of a few tens kV/cm; this permits attaining high charge gains, with avalanche sizes of $10^5$ electrons in a single electrode and up to $10^7$ electrons in cascaded-THGEM configurations, in a variety of gases, including noble ones [2],[13]-[16].

The THCOBRA development aims at combining the advantages of both the MHSP and the THGEM. In this work we describe recent results obtained with the THCOBRA operating in Ne, Ar and P10 (Ar 90% -10 %CH$_4$).

## 2. The THCOBRA

The THCOBRAs investigated (figure 1), with an active area of 15*15 mm$^2$, are made out of a t = 0.4 mm thick G10 plate, covered with copper on both sides. The holes in the G10 plate have d = 0.3 mm diameter with a h = 0.1 mm rim, etched around each hole on both faces. The latter reduces discharge probability, leading to ~ 10-fold higher attainable gains [3][13].

While one of the faces has a continuous electrode, the other one has a pattern etched in the Cu layer, of two independent electrodes: a circular, 0.1 mm wide, cathode surrounding each hole and a 0.3 mm wide anode strip winding between the holes (figure 1). The pattern is similar to that of the COBRA-MHSP [17], but with over five-fold expanded dimensions. There is a bare (uncoated) gap of 0.1 mm between anode and cathode.

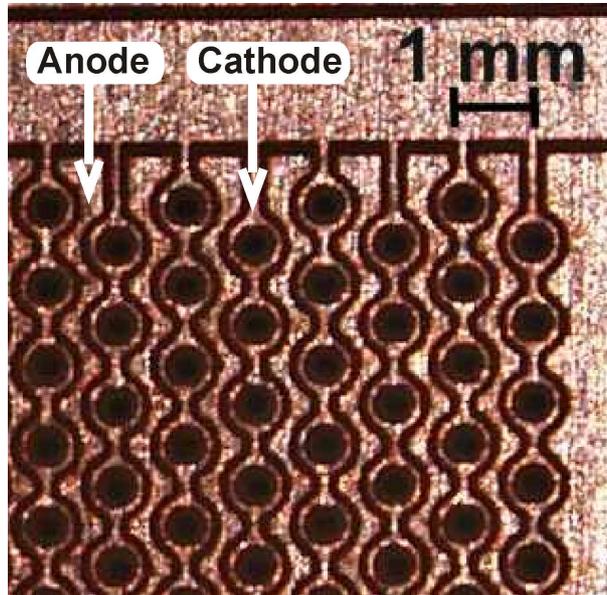

Figure 1: Photography of the THCOBRA. The 0.3 mm diameter holes are drilled in a 0.4 mm thick G10 plate. A 0.1 mm wide, rim is etched around each hole for increased stability. 0.1 mm wide, interconnected cathodes surround each hole. A snake-like anode strips winds between the holes, at a 0.1 mm distance from the circular cathodes.

The THCOBRA has an operation principle (figure 2) similar to that of the MHSP, with two independent multiplication regions, as described in detail in refs.[1][5] . Electrons deposited in the drift region are collected by the dipole field into the holes; hole-multiplication is controlled by the voltage difference between the top and the circular-cathode electrodes, $V_{C-T}$, under a field of a few tens of kV/cm, similarly to the THGEM [18]. The avalanche electrons are extracted from the hole and focused onto the strip-anodes, where a second multiplication occurs, controlled by the voltage between anode and circular-cathode, $V_{A-C}$. Note that the total voltage across the THCOBRA, $V_{TOTAL}$, is the sum of the voltage difference in the holes ($V_{C-T}$) and between anode- and cathode-strips ($V_{A-C}$).

– 2 –

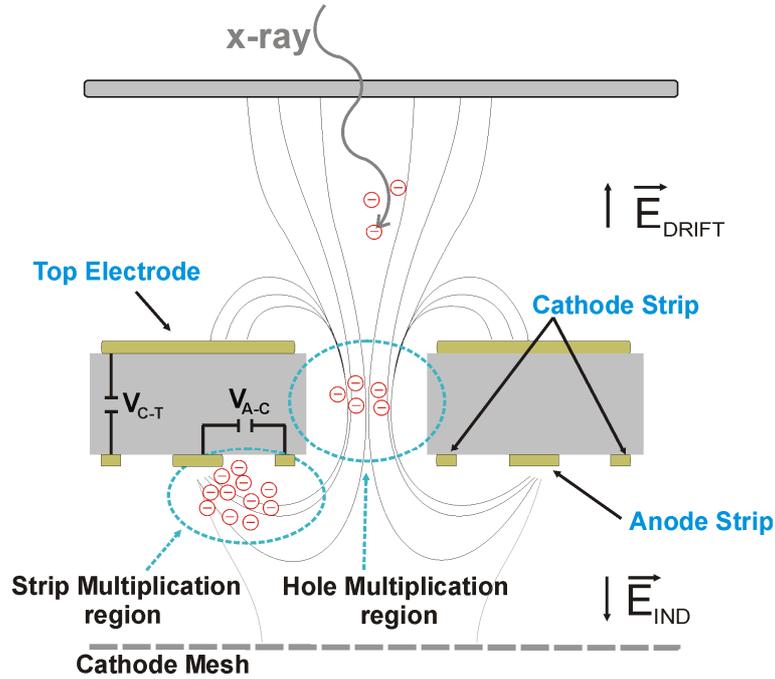

Figure 2: Schematic representation of the THCOBRA operating principle.

To prove that indeed a two-step multiplication can occur within the THCOBRA multiplier, we used field and charge simulations. For values of $V_{A-C} = 300$ V, as shown in figure 3, the electric field intensity, calculated with Maxwell software package [19], in the region between anodes and cathodes, reaches values of several tents of kV/cm , being particularly intense in the vicinity of the anodes. An electron extracted from the holes of the THCOBRA drifts trough the region between anodes and cathodes under the influence of a large electric field, experiencing additional gas multiplication, until the collection at the anode strips.

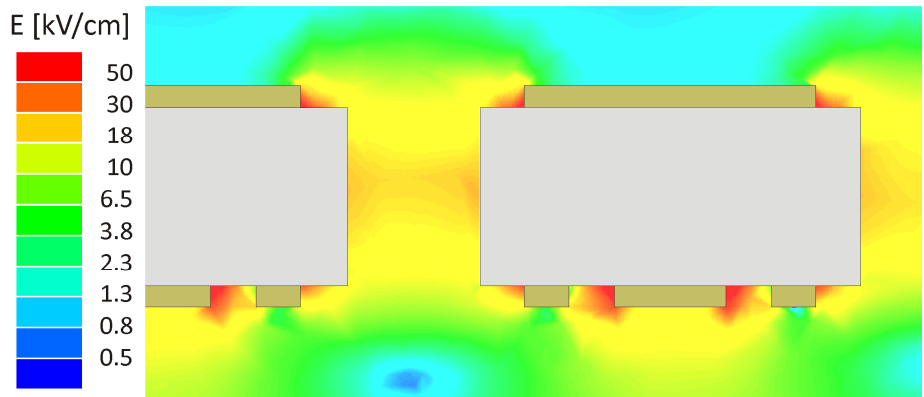

Figure 3: Electric field intensity on the THCOBRA, calculated by the Maxwell software package [19] for $V_{C-T} = 1180$ V, $V_{A-C} = 300$ V, $E_{DRIFT} = 0.1$ kV/cm and $E_{IND} = 4.0$ kV/cm.

The charge multiplication in the vicinity of the anode strips was evaluated with the GARFIELD software package [20]. The path of an electron created in the second, strip multiplication region, and collected at the anodes, was simulated for $V_{A-C} = 60$ V and $V_{A-C} = 280$ V (figure 4). The results indicate that the increase in $V_{A-C}$ causes additional charge multiplication in the region near the anodes, which does not occur at $V_{A-C} = 60$ V.

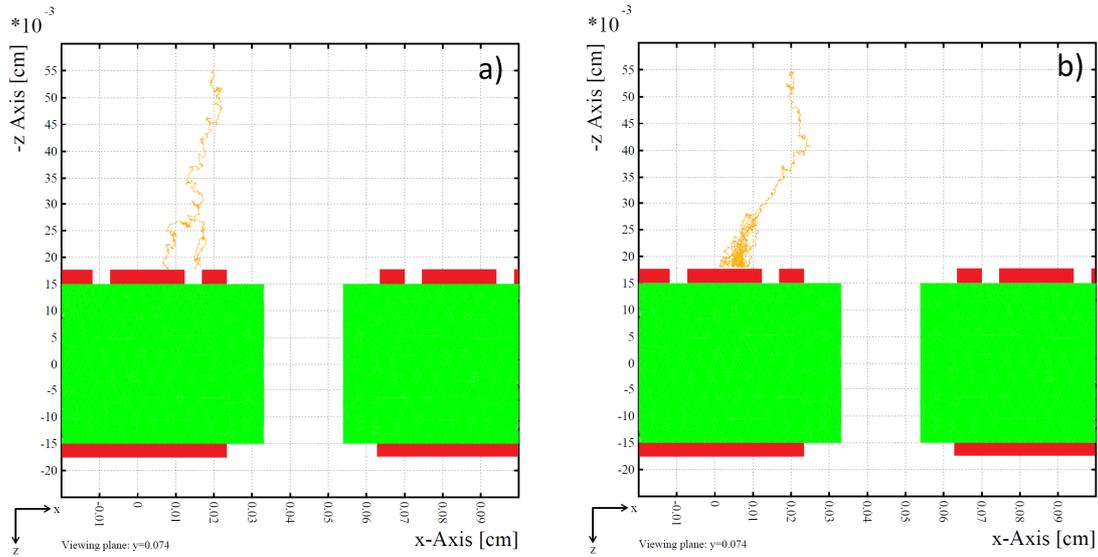

Figure 4: Simulated path and multiplication of an electron created on the strip side of the THCOBRA (and not crossing the holes) for $V_{A-C} = 60$ V (a) and $V_{A-C} = 280$ V (b). Additional charge multiplication occurs for higher values of $V_{A-C}$. Simulations were calculated for $V_{C-T} = 180$ V, $E_{DRIFT} = 0.5$ kV/cm and $E_{IND} = 8.0$ kV/cm.

## 3. EXPERIMENTAL METHODS AND SETUP

The THCOBRA has been investigated within a stainless steel vessel, evacuated down to $10^{-6}$ mbar prior to gas filling. The detector was operated at atmospheric pressure in Ar, and P10 and at 1.7 bar in Ne. During the measurements the detector was operated in a sealed mode; the gas was continuously purified by its circulation through non-evaporable getters, SAES St707, at 200 ºC [7]. The electrical connections were made of UHV-compatible MACOR feedthroughs. All electrodes were independently polarized using CAEN N471A power supplies with current limitation (100 nA). The 3 mm wide induction/extraction region below the THCOBRA (figure 2) was delimited by a metallic grid (80 μm diameter wires, 900 μm spacing) set at ground potential. During the measurements the detector was operated both in pulse-counting and in current (DC) modes.

In pulse-counting mode the detector was equipped with an aluminized Mylar window (25 μm thick) glued to the vessel with a low out-gassing conductive epoxy (Tra-Con 2116). The distance between the detector window and the top electrode of the THCOBRA, defining the drift region, was 27 mm; charges were deposited in the drift region by 22.1 keV x-rays from $^{109}$Cd. The final avalanche-charge per event was recorded from the anodes with a Canberra 2006 preamplifier (sensitivity set to 47 mV/MeV) followed by a Tennelec TC 243 linear amplifier (8 μs shaping time) and a Nucleus PCA 2 multichannel analyzer. The electronic chain sensitivity was calibrated by the injection of a known charge into the preamplifier input.

The current-mode operation was investigated with UV-photons. The Mylar window was replaced by a 5 mm thick UV-transparent quartz window; a second quartz plate, coated with a semi-transparent CsI photocathode, was placed 10 mm above the top electrode of the THCOBRA. The latter, irradiated with a Hg(Ar) UV- lamp (model Oriel 6035), was used as a source of photoelectrons. The primary photocurrent extracted from the CsI photocathode, $I_{PC0}$, depended on the lamp intensity and on the electron extraction field at the surface of the photocathode; a field of 0.1 kV/cm assured good primary electron focusing into the holes and stable primary current. $I_{PC0}$ was recorded prior to each measurement, without charge multiplication in the detector - in order to discard ion back-flow contributions [9]. The maximum values of the currents on the electrodes of the detector were always kept below 100 nA using absorbers placed between the UV lamp and the detector window when required. The final current collected at the anodes of the THCOBRA, $I_A$, was obtained by measuring the voltage drop across a resistor connected in series with the CAEN N471A power supply; the primary current $I_{PC0}$ was measured using a Keithley 610 C electrometer. The values of these currents were used to calculate the charge gain of the detector, corresponding to the ratio $I_A/I_{PC0}$.

## 4. RESULTS AND DISCUSSION

### 4.1 Charge Gain Measurements

The charge gains obtained in 1 bar Ar and P10, and in 1.7 bar Ne are presented in figure 5 as a function of the total voltage ($V_{TOTAL}=V_{A-C}+V_{C-T}$) applied to the THCOBRA electrodes.

The results in Ne were obtained operating the detector in current mode. $V_{C-T}$ was first raised to 340V while $V_{A-C}$ was kept at 0 V by raising simultaneously the potentials of the THCOBRA anodes and cathodes. Following that, $V_{C-T}$ was kept at 340 V and $V_{A-C}$ was raised up to 140 V; this was achieved by keeping the cathode voltage constant while increasing the anode voltage - resulting in a total voltage across the THCOBRA of 480 V. For values of $V_{A-C} = 0$ V (corresponding to the situation where only $V_{C-T}$ was being increased) the current on the anode was zero and the total gain of the detector was obtained by dividing the cathode current by the primary current, $I_{PC0}$. The increase of $V_{A-C}$ led to an increase in the anode current and to a rapid decrease of the current measured on the cathode; the latter dropped to zero for $V_{A-C} > 80$ V.

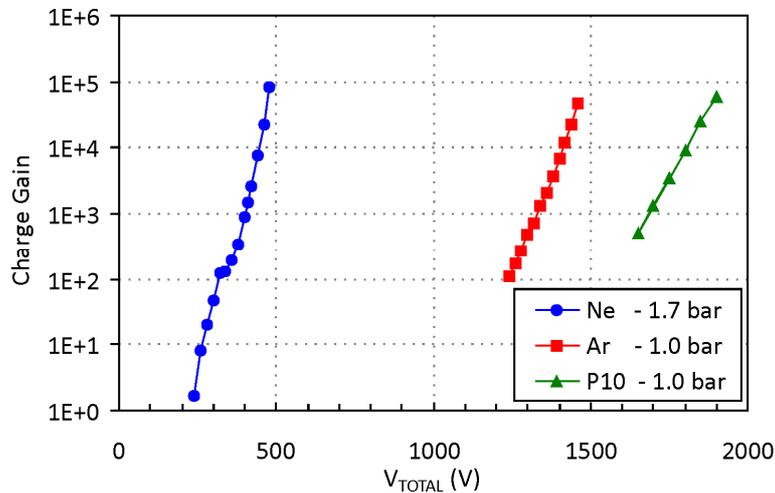

Figure 5: Charge gain obtained with the THCOBRA operating in Ne, Ar, and P10 as a function of the total voltage applied to the T-MHSP. The results in Ar and P10 were measured in pulse-counting mode, with 22.1 keV x-rays; that in Ne were recorded with UV-photons, in current mode.

The results presented in figure 5 for Ar and P10 were taken in pulse-counting mode, using $^{109}$Cd x-rays to induce the primary charge, collecting the final charge at the anodes after multiplication. The voltage across the holes, $V_{C-T}$, was 1180 V for the measurements in Ar and 1475 V for the ones in P10. The maximum gains obtained in these gases were similar: $5*10^4$ in Ar and $6*10^4$ in P10, at respective total voltages of 1460 V and 1900 V.

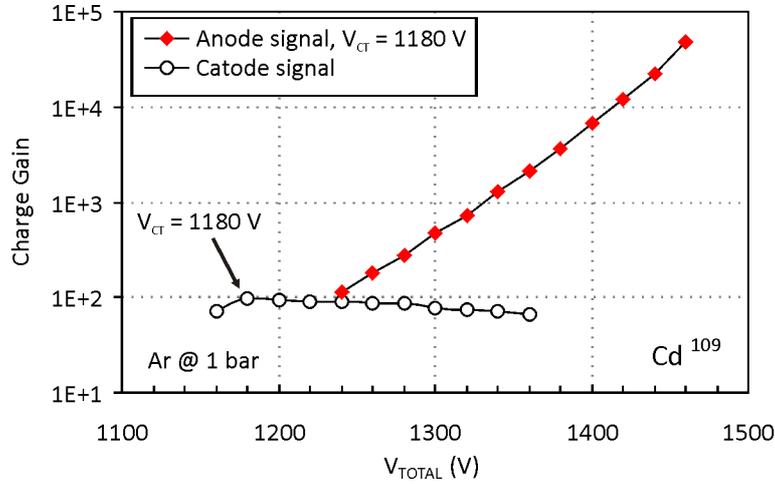

Figure 6: Charge gain measured simultaneously at the anodes and cathodes of the THCOBRA. The first point in the anode-signal curve was obtained for $V_{A-C} = 60$ V. Below this value, the signal recorded in the pre-amplifier connected to the anode was within the noise, i.e. absolute gain less than 70.

In order to estimate the contribution to the total gain of the charge multiplication in the region between the anodes and the circular-cathodes, the charge gain was independently evaluated on these two electrodes, operating the detector in pulse-counting mode. For this the charge collected at the circular-cathode was measured with a second Canberra 2006 pre-amplifier (followed by a Tennelec TC 243 amplifier); both anode and cathode signals were recorded simultaneously. The results presented in figure 6, in argon at atmospheric pressure were recorded by increasing $V_{C-T}$ to the maximum value, while keeping $V_{A-C} = 0$ V (the first two points in cathode signal data), resulting in a null electric field between the anodes and cathodes. Once $V_{C-T}$ reached its maximum value (1180 V) the potential at the anodes was increased up to a total voltage of 1460 V. The metallic mesh bellow the THCOBRA was grounded to ensure that all the negative charge was collected at the THCOBRA electrodes.

The operation with $V_{A-C} = 0$ V represents maximum gains, measured at the cathode of the THCOBRA, of approximately 100 for $V_{C-T} = 1180$ V (figure 6). At this point the signal on the anode electrodes was below the detection threshold, i.e. gain 70.

The increase in $V_{A-C}$ from 0 V to 280 V led to a total gain of $5*10^4$, measured at the anode electrode, representing an increase by a factor of 500 relatively to the charge gain measured at the cathodes for $V_{A-C} = 0$V. This increase is partly due to an improvement of the charge collection on the anode strips but mostly to the additional charge multiplication along the electron path from the holes to the anode strips and in the region around these strips (figure 3).

In figure 7 we present a pulse-height distribution recorded in P10, irradiating the detector with $^{109}$Cd x-rays. The distribution features the Ag kα and kβ x-rays, the Cu k-fluorescence lines from the copper electrodes and the respective escape peaks. An energy resolution of 12.2% FWHM was measured for the 22.1 keV energy x-rays and 19.2% for the 8 keV Cu fluorescence line.

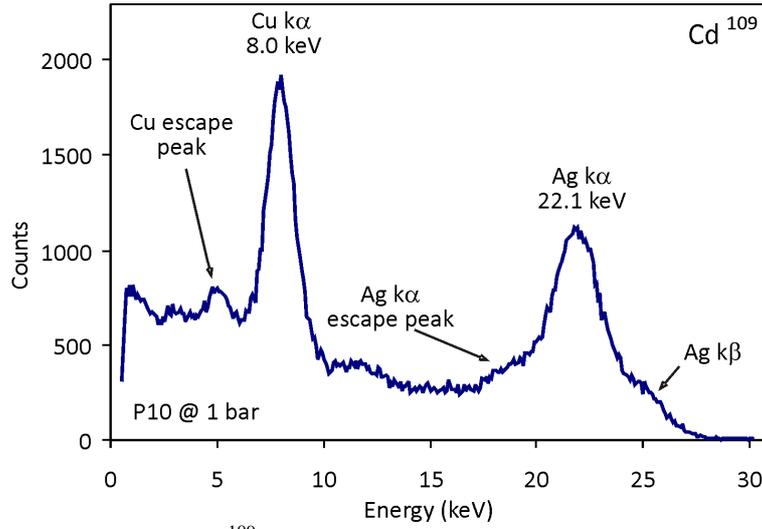

Figure 7: Energy distribution of a $^{109}$Cd radioactive source. Respective resolutions of 12.2% and 19.2% (FWHM) were measured for the 22.1 and 8 keV lines in P10, at 1 bar, at charge gain of $10^4$. The energy resolution values were obtained by fitting a Gaussian curve superimposed on a linear background to the region of interest around the peaks.

### 4.2 Single Photo-electron Measurements

The relatively high charge gains recorded with the THCOBRA could permit, with proper electronics, the detection of single photoelectrons. The response to single photoelectrons was investigated with a THCOBRA coupled to a semitransparent CsI photocathode irradiated with a collimated UV beam; the detector was operated in pulse-counting mode. The single photo-electron pulse-height distributions are presented in figure 8 for different $V_{TOTAL}$ values.

The average charge gain for each distribution, Q, presented in figure 8, was obtained by fitting the single-photoelectron spectra to a normalized Polya distribution [21][22],

$$P(q) = \frac{1}{Q} e^{\frac{-q}{Q}}$$

were $q$ is the individual gain of each avalanche. The average charge gains reached in single-photoelectron conditions, e.g. $2 \times 10^4$ at $V_{TOTAL} = 1900$ V, are similar to those obtained with x-rays (figure 5). As expected, one can reach higher biasing voltages in the single photo-electron conditions, due to higher onset of the Raether limit.

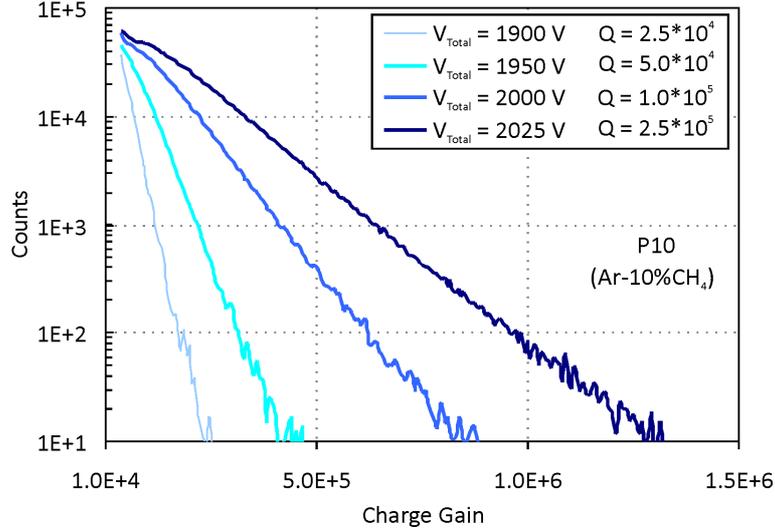

Figure 8: Single-photoelectron spectra obtained with the THCOBRA in P10 at 1 bar.

## 5. CONCLUSIONS AND FUTURE PROSPECTS

We described a novel THGEM-like thick hole-multiplier, with additional patterned electrodes: circular cathode strips around the holes and anode-strips winding between them. Charges are collected and multiplied in the holes, with additional multiplication of the avalanche electrons on the region between the holes and the anode strips. Some of the THCOBRA operation properties were characterized in pure noble gases, Ar and Ne, and in P10. Charge gains of $5*10^4$ and of $6*10^4$ were attained with 22.1 keV x-rays at atmospheric pressure in Ar and P10. Charge gains of $10^5$ were measured in Neon at 1.7 bar with UV photons, in current mode. Significant multiplication was measured on the anode strips, due to the high local fields. An energy resolution of 12.2% was obtained with 22.1 keV kα x-rays in P10.

The results indicate the possibility of reaching reasonable charge gains with a simple-to-produce single-element patterned hole-multiplier. One of the immediate advantages of the THCOBRA is the relatively low voltage required to obtain gains similar to the ones achieved with other thick electron multipliers operating at higher voltages. This feature can be of great interest in applications in dense gases (such as high-pressure and cryogenic detectors). On the other hand the signal collected at the anodes will have a long ion component compared to the electron one only of a THGEM.

An ongoing study [23] is that of ion trapping with THCOBRA. It follows the most successful applications of the MHSP for avalanche ion back-flow (IBF) reduction in electron multipliers, taking advantage of ion trapping by proper polarization of the patterned electrodes [4][9][11].

Further, systematic, investigations are necessary for assessing important parameters in view of different potential applications, among them: gain-stability, counting-rate capabilities, absolute single-photoelectron detection efficiencies, ion blocking capability etc. The design and geometry of the THCOBRA electrodes are subject to optimization by simulations.

## Acknowledgments

This work was supported by projects CERN/FP/109324/2009 and CERN/FP/109283/2009 through FEDER and FCT (Lisbon) programs. A. Breskin is the W.P. Reuther Professor of Research in the Peaceful use of Atomic Energy. F. D. Amaro acknowledges grant SFRH/BD/30318/2006 from FCT (Lisbon). The THCOBRA tested was manufactured by Print Electronics, Israel.